\documentclass[prl,twocolumn,superscriptaddress,preprintnumbers,a4paper,amsmath,amssymb,showpacs,floatfix]{revtex4}
\usepackage{graphicx}
\usepackage{graphics}
\usepackage{bm}
\usepackage{dcolumn}

\newcommand{\crs}{CeRuSn}

\begin{document}

\title{Coexistence of different magnetic moments in CeRuSn probed by polarized neutrons}

\author{K.~Proke\v{s}}
\email{prokes@helmholtz-berlin.de}
\affiliation{Helmholtz-Zentrum
Berlin f\"{u}r Materialien und Energie, Hahn-Meitner Platz 1, EM-AQM, 14109 Berlin, Germany}

\author{S. Hartwig}
\affiliation{Helmholtz-Zentrum
Berlin f\"{u}r Materialien und Energie, Hahn-Meitner Platz 1, EM-AQM, 14109 Berlin, Germany}

\author{A.~Gukasov}
\affiliation{Laboratoire L\'{e}on Brillouin CEA-Saclay - Gif-sur-Yvette 91191, France}

\author{J.A.~Mydosh}
\affiliation{Kamerlingh Onnes Laboratory, Leiden University, 2300 RA Leiden, The Netherlands }

\author{Y.-K. Huang}
\affiliation{Van der Waals-Zeeman Institute, University of Amsterdam, 1018XE Amsterdam, The Netherlands}

\author{O.~Niehaus}
\affiliation{Institut f\"{u}r  Anorganische und Analytische Chemie, Westf\"{a}lische-Wilhelms-Universit\"{a}t M\"{u}nster, 48149 M\"{u}nster, Germany }

\author{R.~P\"{o}ttgen}
\affiliation{Institut f\"{u}r  Anorganische und Analytische Chemie, Westf\"{a}lische-Wilhelms-Universit\"{a}t M\"{u}nster, 48149 M\"{u}nster, Germany }

\date{\today}

\pacs{75.25.-j, 75.30.-m}
\begin{abstract}
We report on the spin densities in \crs~determined at elevated and at low temperatures using polarized neutron diffraction. At 285 K, where the \crs~crystal structure, commensurate with the CeCoAl type, contains two different crystallographic Ce sites, we observe that one Ce site is clearly more susceptible to the applied magnetic field whereas the other is hardly polarizable. This finding clearly documents that distnictly different local environment of the two Ce sites causes the Ce ions to split into magnetic Ce$^{3+}$ and non-magnetic  Ce$^{(4-\delta)+}$ valence states. With lowering the temperature, the crystal structure transforms to a structure incommensurately modulated along the $c$ axis. This leads to new inequivalent crystallographic Ce sites resulting in a re-distribution of spin densities. Our analysis using the simplest structural approximant shows that in this metallic system Ce ions co-exist in different valence states. Localized 4$f$ states that fulfill the third Hund's rule are found to be close to the ideal  Ce$^{3+}$ state (at sites with the largest Ce-Ru interatomic distances) whereas Ce$^{(4-\delta)+}$ valence states are found to be itinerant and situated at Ce sites with much shorter Ce-Ru distances. The similarity to the famous $\gamma$-$\alpha$ transition in elemental cerium is discussed.

\end{abstract}

\maketitle

\section{Introduction}

Ce-based systems are known to exhibit different magnetic states that include the Kondo effect, heavy fermion and superconductivity states, often in coexistence \cite{CRSAF38}. On the other hand, elemental Ce is well known for its $\gamma-\alpha$ transition \cite{CRSAF37}. The exact nature of this transition and the state of the Ce ions in the two phases remains still under debate~\cite{CRSAF35,CRSAF36}.

\crs~ adopts above room temperature a crystal structure with space group $C/2m$ that is related to the monoclinic structure of CeCoAl (the $c$-lattice parameter $\approx$ 5.1 \AA) by a doubling of the cell along the $c$-axis (denoted here as the 2$c$ structure with $c$ $\approx$ 10.2 \AA)~\cite{CRSAF3,CRSAF1,CRSAF5,CRSAF7,CRSAF6}. In \crs~at room temperature, originally single Ce, Ru and Sn crystallographic positions are split into two inequivalent sites (denoted here as Ce1 and Ce2). In the superstructure, each Ce remains coordinated by twelve Ru and Sn atoms. However, the two Ce ions have distictly different nearest Ce-Ru and Ce-Sn neighbor distances, those for Ce1 being significantly shorter. 

\begin{figure}
\includegraphics*[scale=0.3]{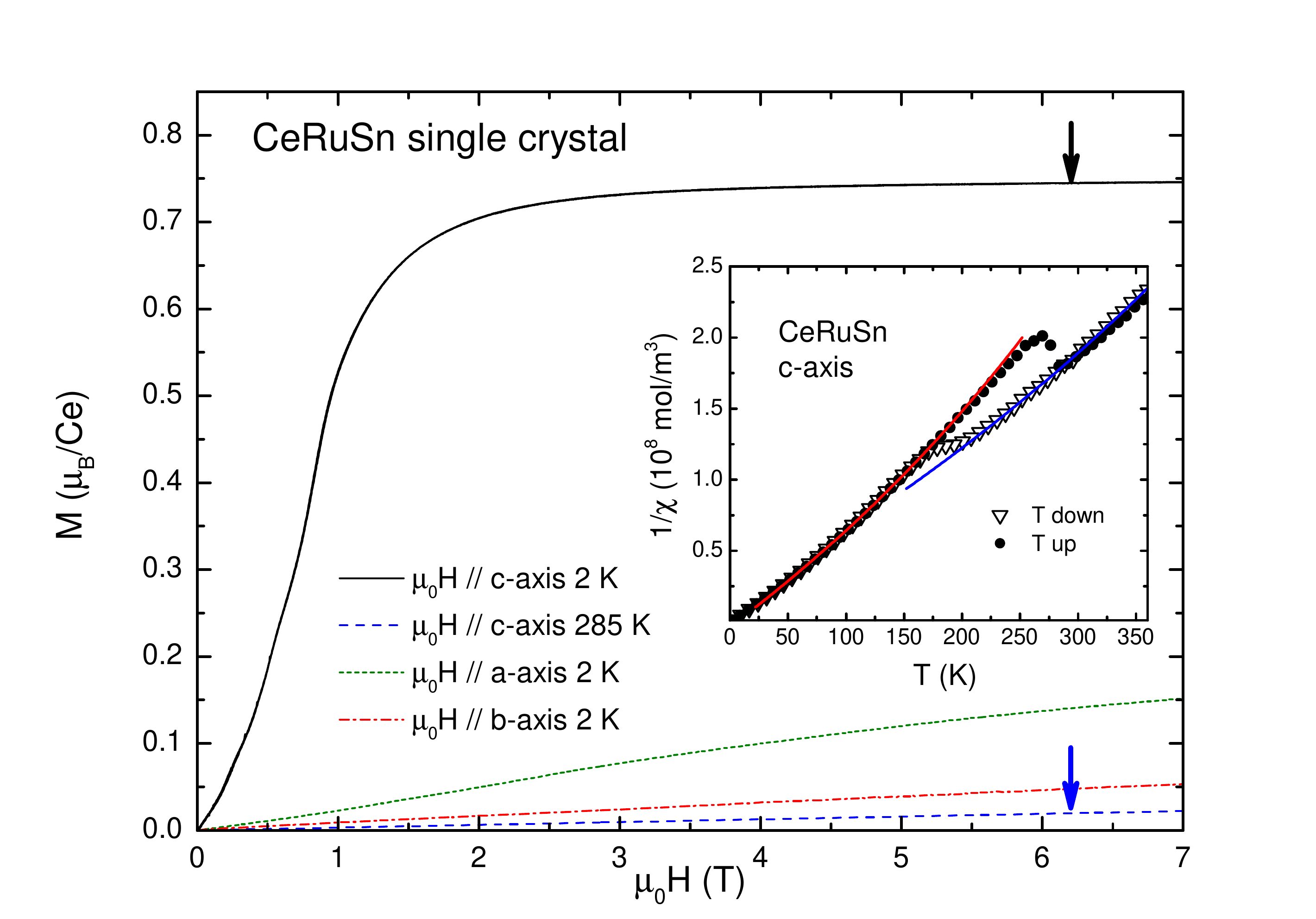}
\caption{(Color online) Field dependence of the \crs~single crystal magnetization measured at 2 K up to 7 T applied along the principal directions and at 285 K along the $c$-axis. Arrows denote positions at which the polarized neutron experiment has been performed. In the inset, the temperature dependence of the inverse magnetic susceptibility 1/$\chi(T)$ = H/M(T) measured upon cooling and warming-up in a field of 1 T applied along the $c$-axis is shown. Best fits to a modified Curie-Weiss law in the range between 250 and 350 K and between 50 and 150 K (both upon cooling) are shown by lines through the experimental points, respectively. Note the different slopes above and below the hysteretic structural transition.} \label{fig1}
\end{figure}

Below room temperature, a hysteretic structural phase transformation occurs in \crs~that is visible from temperature dependencies of various physical properties including the electrical resistivity, thermal expansion and magnetic susceptibility~\cite{CRSAF1,CRSAF7,CRSAF4,CRSAF21,CRSAF2}. Earlier diffraction experiments were interpreted as that the 2$c$ structure is replaced by other structural modifications whose volume fractions are temperature and history dependent. It was shown that in the low-temperature limit, the dominant crystal structure mode appears to be close to a tripling of the basic CeCoAl cell (denoted as the 3$c$ structure $c$ $\approx$ 15.3 \AA). However, single crystal work indicates that this is only an approximation to the real crystal structure. Commensurate Bragg reflections are accompained by incommensurate reflections~\cite{CRSAF23} due to an additional modulation of the structure. The \crs~crystal structure can be thus described in four dimensions using the superspace symmetry ~\cite{CRSAF13}. The average structure is of the CeCoAl type and described within the three-dimensional space group $C2/m$. The fourth dimension describes shifts of Ce, Ru and Sn atoms within the x-z plane according to the modulation vector of $\thickapprox$ 0.35 propagating along the $c$-axis~\cite{CRSAF23}. In fact, the high-temperature crystal structure of \crs~can be viewed as already a modulated structure that is commensurate with the CeCoAl type (leading to two inequivalent sites). 
Ab-initio calculations for this kind of structure suggested that \crs~is close to a magnetic instability and predicted an antiferromagnetic (AF)~\cite{CRSAF5} structure for this material. This has been identified below $T_{N}$ = 2.8 K~\cite{CRSAF1,CRSAF4,CRSAF21,CRSAF23}. It was also predicted theoretically that not all the Ce ions are expected to be magnetically ordered. Ions at the Ce1 site should be non-magnetic whereas those at the Ce2 sites are suposed to be magnetic~\cite{CRSAF3,CRSAF5,CRSAF4}. It has been speculated that Ce1 ions are in the intermediate-valence Ce$^{(4-\delta)+}$ state and Ce2 close to trivalent Ce$^{3+}$. The reported magnetic susceptibility behavior is compatible with this prediction~\cite{CRSAF1,CRSAF7,CRSAF4}. Recently, it has been shown that resonant X-ray scattering spectra on a \crs~single crystal indeed exhibit features compatible with a presence of two different Ce valence states ~\cite{CRSAF22}.  However, no direct microscopic proof for the spatial distribution of spin densities (i.e. showing magnetic and non-magnetic Ce ions) that would be in accord with theoretical prediction exists up to date.
	
	The high-temperature commensurate cell becomes modulated incommensurately at lower temperatures leading to a distribution of atomic sites around their high-temperature positions (or better, around average positions in a CeCoAl type of structure). This, in turn, suggests that a re-distribution of valence states is to be expected. Indeed, the resonant X-ray scattering experiment resolved tiny changes in the spectral weights belonging to 4$f^{1}$ and 4$f^{0}$ states as one goes accross the structural transition. However, it was impossible to conclude whether at low temperatures magnetic and non-magnetic Ce moments still coexist and how they spatially distribute. Our previous unpolarized neutron diffraction experiment has suggested a smooth modulation of moments magnitudes between 0.11 and 0.95 $\mu_B$/Ce~\cite{CRSAF23}. However, at the same time it came to a rather surprising result, namely that the largest moments are found for Ce ions that have the shortest distances to their neighbors. 
	
	In this contribution we report polarized neutron diffraction (PND) experiments using a \crs~single crystal performed at high and low temperatures with magnetic field up to 6.2~T applied close to the $c$-axis in order to determine spin densities in the commensurate and incommensurate states and to identify the magnetically different Ce ion sites. At room temperature we observe a spin distribution that is entirely in agreement with theoretical prediction, i.e. a coexistence of easily polarizable and hardly polarizable Ce sites, where the latter is identifyied as site Ce1.  At low temperatures, assuming the simplest commensurate approximant of the incommensurate structure (the trippling of the original CeCoAl subcell leading to three inequivalent Ce sites), we observe a re-distribution of the spin densities, with a significant moment at all Ce atoms. These moments are not equal and in contrast to a previous study, we observe that the largest magnetic moment resides at position with the largest Ce-Ru nearest neighbour distances.

\section{Experimental}

The details regarding the sample preparation and characterization can be found in Ref. ~\cite{CRSAF23}. The crystal had a shape of a prallelepiped with dimensions of 1x1.5x6 mm$^{3}$. Orientation with the Laue backscattering technique shows that the $b$-axis axis is parallel to the longest crystal dimension.  We have used the same crystal also in magnetic bulk measurements. 

Magnetization curves $M$($T$) and the static magnetic susceptibility $\chi$=$M$/$H$, where $H$ denotes the applied magnetic field along principal directions were measured on the same single crystal oriented by the Laue backscattering technique in the temperature range between 1.8 and 350~K using the Quantum Design 14 T Physical Properties Measurements System (PPMS).

Flipping ratios were collected on the single-detector normal-beam diffratometer 6T2 installed at the ORPH\'{E}E 14 MW reactor of the L\'{e}on Brillouin Laboratory, CEA/CNRS Saclay with the incident wavelength $\lambda$ = 1.40 \AA. The polarization of the incident neutron beam was 97 \%. A magnetic field of 6.2 T has been applied nearly along the $c$-axis with the sample wrapped in an aluminium foil to prevent stresses.

 At both temperatures, due to the magnet opening and the fact that the magnetic field had to be applied along the $c$-axis, only ($hkl$) reflections with l $\leq$ 3  (within the average lattice framework) could be measured within the 0.05 $\leq$ sin $\theta$/$\lambda$ $\leq$ 0.46 range.

\begin{figure}
\includegraphics[scale=0.23]{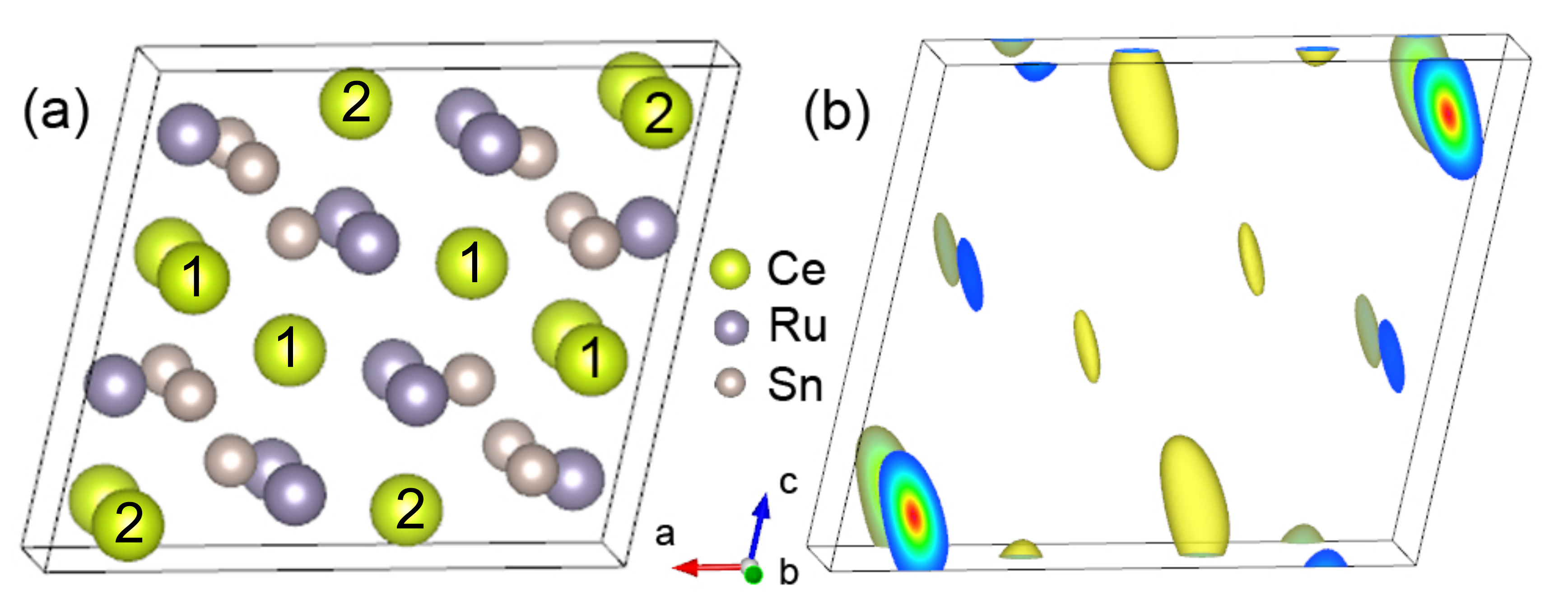}
\caption{A schematic representation of the crystal structure adopted by \crs~at room temperature (a). The atoms are shown as large, intermediate and small circles for Ce, Ru and Sn, respectively and the numbers within the largest circles denote different Ce1 and Ce2 sites with short and large distances to the nearest neighbors, respectively. A color-coded spin density distribution in \crs~obtained from data recorded at 285 K using maximum entropy reconstruction is shown in (b). Note remarkable different densities at the corresponding Ce sites.}
\label{fig2}
\end{figure}

\section{Results}

\subsection{Magnetic bulk measurements}

Previous studies by us and also by other groups show that the temperature dependence of the magnetic susceptibility measured upon cooling and heating exhibits between $\sim$ 170 and $\sim$ 280 K a hysteretic behavior \cite{CRSAF4,CRSAF21} concominant to the crystal structure transformation between the high-temperature (commensurate) and low temperature (incommensurate) crystal structures \cite{CRSAF23}. Outside this region the magnetic susceptibility behaves according to a modified Curie-Weiss (MCW) law. For the measurement performed between 250 and 350 K (above the hysteretic range) and between 50 and 150 K (below the hysteretic range), different parameters are derived. While above the transition we obtain for all three principal directions an effective moment that is reduced with respect to the free Ce$^{3+}$ ion configuration below the transition the effective moment is reduced further by about 20 \%. We associate this decrease with a re-distribution of valence states, namely with an increasing amount of Ce$^{(4-\delta)+}$ states. The best MCW fits for the $c$-axis measurement are shown in the inset of Fig.~\ref{fig1} by full lines.

The response of \crs~ to the applied magnetic field is at all temperatures very anisotropic. Magnetization curves obtained for the field applied along the principal axes of \crs~ at 2 K and along the $c$ axis at 285~K are shown in Fig.~\ref{fig1}. While at high temperature the magnetization for all the directions increases linearly and only very slowly with no saturation tendency, at 2~K two metamagnetic-like transitions are visible for the $c$-axis orientation, with the two remaining orientations showing only slow magnetization increase with no saturation tendency. For the $c$-axis orientation, a magnetization of 0.019(1) $\mu_B$/Ce is reached at 285 K and at 6.2~T. At 2~K and the same field it increases to 0.744(2) $\mu_B$/Ce. Clearly, at higher fields, a field-induced saturated ferromagnetic (F) state is established for the $c$-axis orientation. 

\subsection{Spin density at 285 K}

In order to obtain  reliable information regarding the magnetic spin densities from the polarized neutron experiment, precise information about crystal structure is needed. For the room temperature structure we have utilized parameters determined fom our previous x-ray single crystal diffraction and neutron powder diffraction that are in very good agreement with structural information obtained by other groups ~\cite{CRSAF1,CRSAF2,CRSAF7}. This high-temperature crystal structure of \crs~ projected nearly along the $b$-axis is shown in Fig.~\ref{fig2}(a).

Although the \crs~bulk magnetization at 285 K and at 6.2 T oriented along the $c$-axis is only 0.019(1) $\mu_B$/Ce, this value is sufficiently high to perform a PND experiment and to obtain reliable flipping ratio data. At 285 K, we have collected total flipping ratios from 187 Bragg reflections (85 unique ones).
Magnetic structure factors have been calculated with a help of the Cambridge Crystallography Subroutine Library \cite{CRSAF29} suite programs. Spin densities were determined using the software package PRIMA \cite{CRSAF25} that calculates the most probable distribution that is in agreement with the symmetry of the parent lattice, observed magnetic structure factors and associated errors using the maximum entropy (MAXENT) method~\cite{CRSAF24}. The resulting densities were drawn using the computer code VESTA~\cite{CRSAF26}. 
In Fig.~\ref{fig2}(b) a color-coded spin density distribution as determined from the measured flipping ratios at 285 K is shown as viewed nearly along the $b$-axis.  Comparison with the crystal structure in the same orientation shown in Fig.~\ref{fig2}(a) shows that spin densities in the form of clouds elongated along the $c$-axis (caused by a lesser resolution along the direction of the applied field) are found at places that correspond to both Ce1 and Ce2 atoms. However, the densities are not equal at the two inequivalent sites. The integration corresponding the Ce$^{3+}$ ionic radius of 1.34~\AA for the coordination number 12~\cite{CRSAF27} reveals that while the Ce2 posseses a total magnetic moment of 0.017(3) $\mu_B$, the Ce1 site only 0.004(2) $\mu_B$, i.e. substantially less than at the Ce2 site. These values are listed in Table~\ref{tab:table1}. The average of the two values is smaller than the bulk magnetization of 0.019 $\mu_B$/Ce achieved at 6.2 T suggesting a non-zero contribution from other sites or conduction electrons. Indeed, the integration reveals that also the interatomic space is significantly polarized. The signal intensity, however, does not allow conclusions regarding possible polarization at the Ru/Sn sites.

Another way to treat the experimental data is the direct refinement of the measured flipping ratios. We assume all the magnetic moments to be centered on the given atomic sites. Various atoms are characterized by magnetic form factors f($\textbf{Q}$) that depend on the scattering vector $\textbf{Q}$ and have in general orbital ($\mu _L$) and spin ($\mu _S$) parts, where the total magnetic moment $\mu$  = $\mu _S$ + $\mu _L$.

The cerium magnetic form factor is usually expressed within the dipolar approximation by the formula $f(\textbf{Q}$) = $\langle j_0$ ($\textbf{Q}$) $\rangle + C_2 \langle j_2 (\textbf{Q}) \rangle $, where $C_2$ = $\mu _L  / ( \mu _S  + \mu _L  )$ and $j_i$ is the radial integral for the Ce$^{3+}$ valence state~\cite{CRSAF30}. An equivalent expression can be written down for the Ru magnetic form factor. The best fit allowing magnetic moments on Ce and Ru sites indicates that no significant magnetic moment resides at neither of the two inequivalent Ru sites. By assuming a magnetic moment on the Ce sites only, we could obtain a rather good fit of the experimental data. The best fits yield the moment values listed in Table~\ref{tab:table1}. While it is difficult to conclude anything regarding the coupling of the spin and orbital parts at the Ce1 site, the spin and orbital moments on the Ce2 atoms seem to be clearly parallel to each other with the $C_2$ parameter strongly reduced at 285 K with respect to expected value of 1.33 calculated for both, $\alpha$ and $\gamma$ cerium \cite{CRSAF34}. Thus, the derived magnetic form factor deviates from that of the free ion Ce$^{3+}$ situation by having different coupling of the spin and orbital part. We interpret this as a consequence of 4$f$ electron delocalization in agreement with the suggestions made for  $\alpha$ cerium \cite{CRSAF31}. 

\begin{figure}
\includegraphics[scale=0.27]{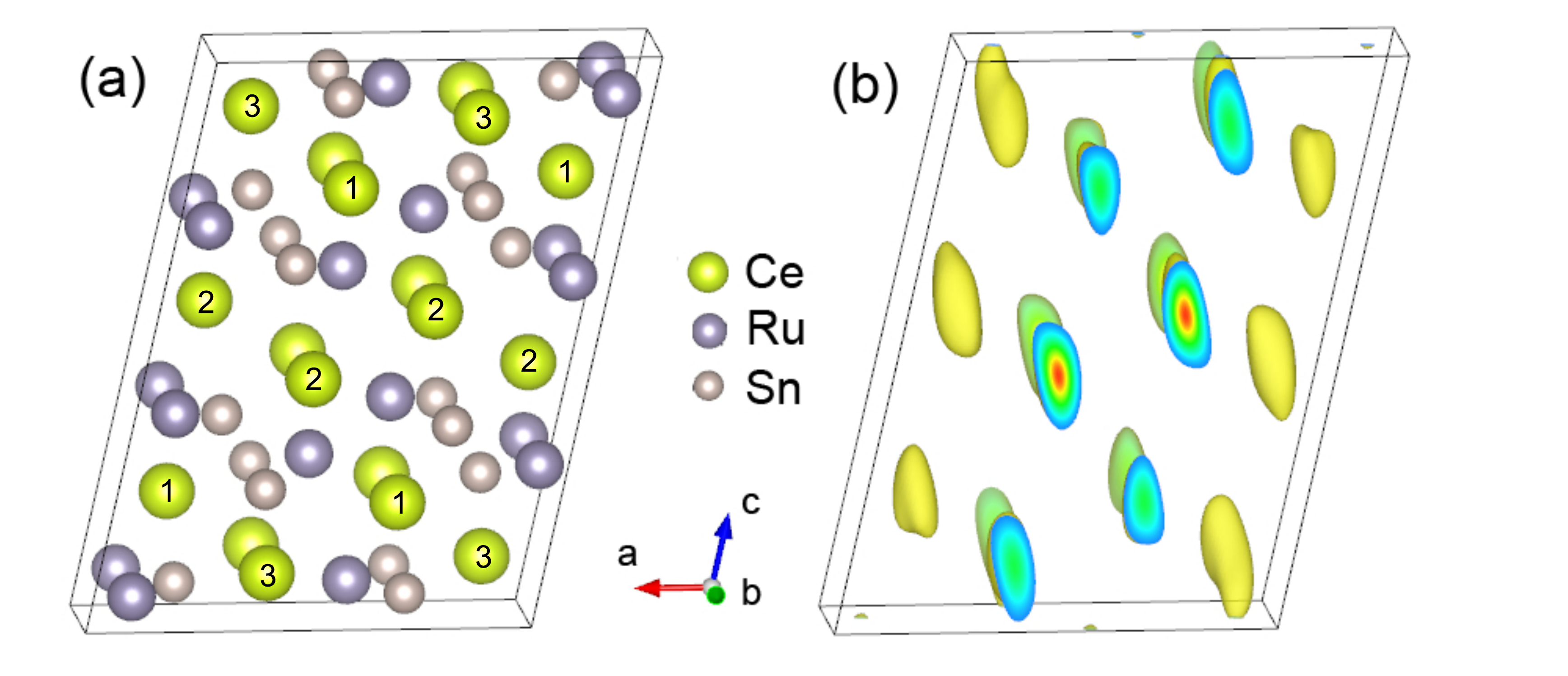}
\caption{A schematic representation of the simplest aproximation to the crystal structure of \crs~at low temperatures (a). The atoms are shown as large, intermediate and small circles for Ce, Ru and Sn, respectively and the numbers within the largest circles denote mutually inequivalent Ce1, Ce2 and Ce3 sites. The Ce3 site has the nearest neighbors at the shortest distances and the Ce2 site at the the largest distance, respectively. A color-coded spin density distribution in \crs~obtained from data recorded at 1.7 K using maximum entropy reconstruction is shown in (b). Note remarkable different densities at corresponding Ce sites.}
\label{fig3}
\end{figure}

\subsection{Spin density at 2 K}

\begin{table}[h]
\caption{\label{tab:table1}Magnetic moment values of a \crs~single crystal
determined from the direct fitting of PND data and by the integration of the spin density maps obtained by
a maximum-entropy method. The PND experiment was carried out at $T =
285$~K and 1.7 K, respectively, in a magnetic field $B \parallel c$ of 6.2 T. We
assumed the Ce moment to have both spin $\mu _S$ and orbital
$\mu _L$ part. The parameter $C_2$ = $\mu _L$ / $\mu_{tot}$ is listed
as well. Bulk magnetization measurements lead magnetization values of 0.019(1) and 0.744(2) $\mu_B$/Ce at 285~K and 2 K, respectively.}

\begin{tabular}{lccccccc}
  \hline
  \hline
T = 285 K  &  &  & & \\
Method  &  & DIRECT & & & MAXENT  \\
\hline
  Site  &  $\mu _S$ & $\mu _L$ & $\mu_{tot}$ & $C_2$ & $\mu_{tot}$ \\
  Ce1 &  -0.006&0.009(7)&0.003(1)&3.0&0.004(2)\\
  Ce2 &  0.009& 0.021(6)&0.030(1)&0.7&0.017(3)\\
    \hline
T = 1.7 K  &  &  &  & & \\
Method  &  & DIRECT & & & MAXENT  \\
\hline
  Site  &  $\mu _S$ & $\mu _L$ & $\mu_{tot}$ & $C_2$ & $\mu_{tot}$ \\
    Ce1 &  0.2&0.4(2)&0.60(3)&0.67&0.58(2) \\
      Ce2 &  -0.2 & 0.9(2)&0.72(4)&1.20&0.82(3)\\
  Ce3 &  0.3&0.1(3)&0.43(4)&0.35&0.42(2) \\
  \hline
    \hline
    \end{tabular}
\end{table}

As mentioned above, below room temperature the commensurate crystal structure becomes incommensurate. Two sets of structural Bragg reflections are observed at 2 K for \crs. The main reflections that correspond to the average CeCoAl type structure and incommensurate ones that can be described by a propagation vector  $q_{nuc} = (0~0~0.35)$~\cite{CRSAF23}. For the polarized data refinement at low temperatures we could use structural information obtained on the same crystal from previous unpolarized neutron diffraction work ~\cite{CRSAF23}. However, due to limitations of the available computer codes~\cite{CRSAF24}, we had to approximate the incommensurate crystal structure by a commensurate one. Since the modulation is close to a trippling of the original CeCoAl structure along the $c$-axis, we have used the three times larger $c$-axis parameter. The positional parameters were taken from literature \cite{CRSAF7,CRSAF2,CRSAF28}. This leads to three independent Ce, Ru and Sn sites. As in the case of the high temperature structure, Ce atoms have distinctly different Ce-Ru and Ce-Sn neighbor distances, those for the Ce3 site being the shortest and the Ce2 the longest. The corresponding structural approximant is shown in Fig.~\ref{fig3}a.

Spin densities were reconstructed from magnetic structure factors calculated from 173 flipping ratios (92 unique ones). These were collected in the same angular range as at high temperatures. The resulting color-coded spin density distribution of \crs~at 1.7 K, projected nearly along the $b$-axis, is shown in Fig.~\ref{fig3}(b). At the first glance, it is apparent that the density clouds correspond very well to positions of Ce ions. The density values are, however, much larger than those at high temperature and different between the sites. Integration of the map in three dimensions, corresponding the the Ce$^{3+}$ ionic radius of 1.34 \AA, reveals values of 0.58(2), 0.82(3) and 0.42(2) $\mu_B$ for the Ce1, Ce2 and Ce3 site, respectivelly. We see that all three sites carry a significant magnetic moment that is, however, much smaller than the Ce free-ion moment of 2.14 $\mu_B$. Also in this case we observe that the average of the three magnetic moments is somewhat smaller than the bulk magnetization, indicating significant polarization at other crystallographic sites and/or intersticial regions. In addition, we observe that in contrast to our previous unpolarized study, the largest magnetic moment is obtained for the site with the largest nearest-neighbor distances.

The second method, a direct refinement to the flipping ratios \cite{CRSAF29}, in which we assume all the magnetic moments to be centered on the Ce atomic sites and described by the Ce$^{3+}$ form factor yields values given in Table~\ref{tab:table1}. Clearly, the agreement between magnetic moments determined from the two methods is very good. The orbital ($\mu _L$) and spin ($\mu _S$) parts were refined independently for the three Ce sites, leading to different $C_2$ parameters that are listed in Table~\ref{tab:table1} as well. An interesting point is that we observe distictly different $C_2$ parameters for the three sites.  At first, all the $C_2$ parameters are strongly reduced for the Ce$^{3+}$ state. Second, there is a clear relation between the $\mu_{tot}$ and $C_2$ on one side and the nearest Ce-Ru distances on the other. The larger the distances, the more develop the magnetic moment and the higher are the $C_2$ value. This is a direct consequence of different spin and orbital parts for the three sites. 

\section{Discussion and Conclusions}

Normally, for systems with less than half-filled $f$-electron shells it is expected that the spin-orbit coupling aligns spin part antiparallel to the orbital part. This effect is known as the third Hund's rule and while for the Ce2 site that has Ru/Sn neibhbours at the largest distances, the rule is fullfilled, the two components are found to be parallel for the Ce1 and Ce3 sites. From the Table~\ref{tab:table1} it is seen that the magnetic moment on the Ce2 site is dominated by the orbital part in contrast to the Ce3 site where the orbital part is considerably quenched and the spin part dominate. These values and their ratio were previously associated with the degree of delocalization in elemental cerium \cite{CRSAF31}. Localized $f$ electrons are supposed to be dominated by the orbital part with a magnetic form factor close to the  Ce$^{3+}$ state. Itinerant 4$f$ states should be dominated by the spin part with the orbital contribution supressed, having the $C_2$ parameter small. In that case it can be shown that the third Hund's rule is violated \cite{CRSAF31}. We see, that within \crs~at low temperatures there are, depending on the Ce atoms, 4$f$ states which behave as localized electrons and states that are itinerant. 

The current observation is also able to shed more light on the mechanism leading to the structural transition itself. Recent resonance X-ray scattering observations show across the structural transition an increase of the 4$f^{0}$ spectral weight \cite{CRSAF22}. Such an inrease is also observed for the $\gamma \rightarrow \alpha$ transition in elemental cerium \cite{CRSAF35} where a 15 \% volume reduction is observed \cite{CRSAF32}. It has been speculated that the mechanism leading to a significant volume collapse in \crs~(with respect to the average CeCoAl type structure) is of a similar type. Two scenarios were put forward in order to explain the isostructural transition in cerium: (i) the Kondo volume collapse where a drastic change in the effective hybridization and thus Kondo temperature occurs and (ii) the orbital-selective Mott transition, in which the hoping between the $f$ orbitals leads to an itinerant $f$ electrons in the $\alpha$ state. Recently, it has been shown experimentally \cite{CRSAF33} and theoretically \cite{CRSAF34} that the magnetic form factor in $\alpha$-Ce is of the Ce$^{3+}$ type discarding the latter model. Accordingly, a hybridization between the localized 4$f$ and the $spd$ electrons leading to a different Kondo energy scales in $\gamma$ and $\alpha$-Ce is identified to be the mechanism behind the volume reduction. The situation in \crs~is similar only to a certain extend. At high temperatures we deal with Ce states that are rather delocalized and not exactly in the Ce$^{3+}$ state. This is also reflected in a reduced  $C_2$ parameter for the Ce2 site that is very similar to this parameter calculated in early $ab-initio$ calculations for $\alpha$-Ce \cite{CRSAF31}. At low temperatures we deal with two thirds of Ce sites where the 4$f$ electrons are delocalized and in a Ce$^{(4-\delta)+}$ valence state. One third of 4$f$ electrons, in contrast, seems to be more localized than at high temperatures.

This means that we must consider two competing mechanisms - one analogous to the Kondo volume collaps mechanism as in elemental cerium, where the 4$f$ electrons are in both $\gamma$ and $\alpha$ states localized and the other that goes just in the opposite direction and leads to more localized states at low temperatures. One can speculate at this moment about the influence of thermal vibrations (phonons) that were identified recently to play an important role in stabilization of the more 4$f$ localized $\gamma$ cerium \cite{CRSAF36}. An inelastic neutron experiment focused on the temperature development of phonon spectra should be able to clarify this.

\acknowledgments We acknowledge the Laboratoire  L\'{e}on Brillouin, CEA Saclay for the allocated beamtime, HZB for financial support and R. Feyerherm for valuable discussion. 


\end{document}